\documentclass[aps,pre,twocolumn,superscriptaddress,showpacs,showkeys]{revtex4}
\usepackage{epsfig}
\usepackage{graphicx}
\usepackage{amsmath}
\usepackage[english]{babel}

\begin{document}

\title{Critical behaviors and universality classes of percolation phase transitions on two-dimensional square lattice}
\author{Yong Zhu}
\author{Ziqing Yang}
\author{Xin Zhang}
\author{Xiaosong Chen}
\affiliation{State Key Laboratory of Theoretical Physics, Institute of Theoretical Physics, Chinese Academy of Sciences, P.O. Box 2735, Beijing 100190, China}

\date{\today}
\begin{abstract}
We have investigated both site and bond percolation on two dimensional lattice under the random rule and the product rule respectively. With the random rule, sites or bonds are added randomly into the lattice. From two candidates picked randomly, the site or bond with the smaller size product of two connected clusters is added when the product rule is taken. Not only the size of the largest cluster but also its size jump are studied to characterize the universality class of percolation. The finite-size scaling forms of giant cluster size and size jump are proposed and used to determine the critical exponents of percolation from Monte Carlo data. It is found that the critical exponents of both size and size jump in random site percolation are equal to that in random bond percolation. With the random rule, site and bond percolation belong to the same universality class. We obtain the critical exponents of the site percolation under the product rule, which are different from that of both random percolation and the bond percolation under the product rule. The universality class of site percolation differs different from that of bond percolation when the product rule is used.
\end{abstract}

\maketitle

\section{Introduction}
\label{introduction} Traditionally, percolation phase transitions are considered to be continuous in various networks. However, Achlioptas et al.~\cite{Achlioptas} concluded that the percolation transition in random network is discontinuous under the Achlioptas process (AP) in which the growth of large clusters is suppressed. During the evolution of a network under AP, edges are added into the network under product rule that the edge with minimum product of cluster sizes is connected from two randomly picked unoccupied edges. The percolation transitions under product rule in two-dimensional square lattice~\cite{ziff09,ziff10,Rad10} and scale-free networks~\cite{Rad10,ChoP09,Rad09} are also investigated and argued to be discontinuous. Later, a series of investigations~\cite{Costa10,Lee11,Grassberger11,Riordan11,Liu11,Fan12} showed that different percolation transitions under product rule are actually continuous. In Ref.~\cite{Costa10},  Costa et al.~\cite{Costa10} studied a model with stronger suppression of large clusters and its percolation transition is continuous. Lee et al.~\cite{Lee11} concluded that the explosive percolation is continuous by studying cluster size distribution. Grassberger et al.~\cite{Grassberger11} found that the explosive transition is still continuous but with unusual finite-size behavior. In the works of Liu et al.~\cite{Liu11} and Fan et al.~\cite{Fan12}, a generalized Achlioptas process (GAP) was introduced to investigate the percolation transitions of two-dimensional lattice network and Erd\"os-R\'enyi (ER) network \cite{ER} respectively. In the generalized Achlioptas process, the edge with minimum product of cluster sizes is connected with a probability $p$ from two randomly picked unoccupied edges. When $p=1/2$, GAP recovers to a random process. With $p=1$, GAP becomes AP. The percolation transitions of two-dimensional lattice network and ER network under GAP keep to be continuous. Their critical exponents and and therefore universality classes depend on the probability parameter $p$. With a rigorous mathematical proof, Riordan and Warnke~\cite{Riordan11} concluded that the percolation transitions under all Achlioptas processes are continuous.

The site percolation on a two-dimensional square lattice under product rule was first investigated by Choi et al.~\cite{Choi11}. They claimed that the percolation is discontinuous from the non-vanishing hysteresis between the directed and reverse process for the fraction of the sites belonging to the largest cluster. However, Bastas et al.~\cite{Bastas11} demonstrated that the hysteresis phenomena will disappear in the thermodynamic limit. Using the finite-size behaviors of the largest cluster and its standard deviation, they concluded that the explosive percolation transition is continuous but belongs to different universality class. To clarify this controversy, further investigations are needed.

The notation of a universality class is a basic tenet in the physics of critical phenomena. Within a universality class, the universal quantities such as critical exponents and scaling functions are independent of microscopic details. It is accepted traditionally that a universality class is characterized by the spatial dimensionality of the system and by the number of the components of the order parameter. (See, e.g., the review article \cite{PAH91}.) On a two-dimensional lattice, the random bond percolation has the same critical exponents as the random site percolation. The random bond and site percolation belong to the same universality class. For the bond percolation under GAP with $p > 1/2$, their critical exponents are different from that of random bond percolation and depend on the probability parameter $p$~\cite{Liu11}.  The universality classes of bond percolation under GAP are different from that of random bond percolation~\cite{Liu11}. It is of interest to investigate the critical behaviors and universality class of site percolation in relation to the Achlioptas process.

In this paper, we study the critical behaviors of site percolation under AP in two-dimensional square lattice. We investigate the sizes of giant clusters in the lattice. From the finite-size scaling behaviors of giant clusters, we can determine the percolation transition point and the corresponding critical exponents. Both the critical exponents of the sizes of giant clusters and their size gaps during evolution~\cite{Lee11,Nagler11,Fan13} are calculated.

Our paper is organized as following. In Section 2, we present the finite-size scaling behaviors of giant clusters near the phase transition point of percolation. The finite-size scaling behaviors of giant clusters for site percolation under random and product rules on two-dimensional lattice are studied in Sections 3 and 4 respectively. The universality classes of bond and site percolation under random and product rules are discussed in Section 5.

\section{Critical behaviors of percolation phase transition}
The percolation phase transition in a network with $N=L^d$ nodes is indicated by the appearance of a giant cluster whose size becomes comparable with $N$.
The size $S_1$ of the giant cluster is taken as the order-parameter of percolation phase transition. Near a critical point, the size $S_2$ of the second largest cluster in the network becomes comparable with $N$ also. Both $S_1$ and $S_2$ demonstrate critical behavior near the critical point \cite{Liu11,Fan12}.

Following the finite-size scaling of order-parameter~\cite{privman1,privman2}, we anticipate that the reduced sizes of $S_1$ and $S_2$ in a network with $N$ nodes and $N_r$ edges follow the finite-size scaling form as ~\cite{Liu11,Fan12}
\begin{eqnarray}
\label{eq2}
s_1(r,L)\equiv S_1/L^d &=& L^{-\beta/\nu} \, \tilde{s}_1(tL^{1/\nu}),\\
\label{eq2a}
s_2(r,L)\equiv S_2/L^d &=& L^{-\beta/\nu} \, \tilde{s}_2(tL^{1/\nu}),
\end{eqnarray}
where $r=N_r/N$ is the reduced number of edges and $t=(r-r_c)/r_c$ is the deviation from critical point $r_c$. The scaling variable $tL^{1/\nu}$ is related to the size ratio $L/\xi$ of $L$ to the correlation length in bulk $\xi= \xi_0 |t|^{-\nu}$. The finite-size scaling form is supposed to be valid in the asymptotic critical region with $L\gg 1$ and $|t|\ll 1$.

From Eqs.(\ref{eq2}) and (\ref{eq2a}), the finite-size scaling form of size ratio $s_2/s_1$ is obtained as
\begin{equation}
s_2/s_1=\tilde{s}_2(tL^{1/\nu}) /\tilde{s}_1(tL^{1/\nu})\equiv
U(tL^{1/\nu}).  \label{eq5}
\end{equation}

At the critical point $r_c$, the size ratio $s_2/s_1=U(0)$ and becomes independent of system size $L$. The curves $s_2/s_1$ of different $L$ against $r$ have a cross-point, which can be used to determine the critical point $r_c$.

The logarithm of Eq.(\ref{eq2}) is
\begin{equation}
\ln s_1(r,L)=-(\beta / \nu)\ln L +\ln \tilde{s}_1(tL^{1/\nu}).
\label{eq6}
\end{equation}
At the critical point, we have
\begin{equation}
\ln s_1(r_c,L)=-(\beta / \nu)\ln L +\ln \tilde{s}_1(0)\;,
\label{eq7}
\end{equation}
which is a straight line against $\ln L$. We can fix the critical point $r_c$ by the straight line of $\ln s_1(r,L)$ with respect to $\ln L$. From the slope of the straight line, we can determine the critical exponent ratio $\beta/\nu$. By introducing scaling variable $tL^{1/\nu}$ to make curves $s_2/s_1$ of different $L$ collapse into a scaling function $U(tL^{1/\nu})$, the critical exponent $\nu$ can be estimated.

There are critical behaviors also during the evolution of a network. The percolation phase transition is accompanied not only by the appearance of a giant cluster but also by the giant size jump of the largest cluster. In the Refs.\cite{Nagler11} and \cite{Fan13}, the percolation phase transition was investigated by the size jumps of the largest cluster.

In a simulation of network evolution, edges are added under some rule one by one into network. At an evolution step $T$, the number of edges in the network increases from $T-1$ to $T$ and the size of the largest cluster varies correspondingly from $S_1 (T-1)$ to $S_1 (T)$. At this step, the largest cluster has a reduced size jump $\delta_T \equiv \left[S_1 (T)-S_1 (T-1)\right]/N$. The largest reduced size jump of the whole evolution process is
\begin{equation}
\Delta \equiv max \left\{ \delta_1,\delta_2,\delta_3,...\right\}.
\end{equation}

The evolution step, where the largest size jump appears, is denoted as $T_c$. In the $i$-th simulation, we obtain the largest reduced size jump $\Delta^{(i)}$ and the critical reduced evolution step $r_c^{(i)}=T_c^{(i)}/N$. From the results of $M$ simulations, we can calculate following averages
\begin{eqnarray}
\bar{\Delta}(L)&=& \frac{1}{M}\sum_{i=1}^{M} \Delta^{(i)}\;,\\
\bar{r}_c (L) &=& \frac{1}{M}\sum_{i=1}^{M} r_c^{(i)}\;.
\end{eqnarray}

In the bulk limit $L \to \infty$, we suppose that $\bar{r}_c (L)$ approaches its bulk limit $r_c (\infty)$ as
\begin{equation}\label{rc}
\bar{r}_c (L)= r_c (\infty)+a_r L^{-1/\nu_1}.
\end{equation}
The bulk limit $r_c (\infty)$ from the network evolution should be equal to the critical point $r_c$ obtained from the largest cluster.

The character of percolation phase transition can be determined from the finite-size effect of $\bar{\Delta} (L)$. If $\bar{\Delta} (L)$ approaches a non-zero value in the limit $L \to \infty$, the percolation is a discontinuous phase transition. For a continuous percolation phase transition, $\bar{\Delta} (L)$ has a power-law finite-size effect as
\begin{equation}\label{delta}
\bar{\Delta} (L)=a_\Delta L^{-\beta_1}.
\end{equation}

In the $i$-th simulation of network evolution, there are fluctuations $\delta r_c = r_c^{(i)} - \bar{r}_c (L)$ and $\delta \Delta = \Delta^{(i)} - \bar{\Delta}(L)$. Their root mean squares are calculated as
\begin{eqnarray}
\label{eq8}
\chi_r &\equiv& \sqrt{<(\delta r_c)^2>}, \\
\label{eq9}
\chi_\Delta &\equiv& \sqrt{<(\delta \Delta)^2>}.
\end{eqnarray}

Their dependence on network size $L$ is described by the exponents $\nu_2$ and $\beta_2$  as
\begin{eqnarray}
\label{eq10}
\chi_r &\propto& L^{-1/\nu_2}, \\
\label{eq11}
\chi_\Delta &\propto& L^{-\beta_2}.
\end{eqnarray}

In Ref.~\cite{Nagler11}, the asymptotic behavior of $\bar{\Delta} (L)$ was used to judge the continuity of a percolation phase transition. The so-called upper pseudo transition point in Ref.~\cite{Lee11} was estimated from $\bar{r}_c (L)$. The general clique percolation phase transition in random networks are identified by studying the averages and fluctuations of $\Delta$ and $r_c$~\cite{Fan13}.

We will investigate the above critical behaviors of a percolation phase transition, which are characterized by the critical exponents $\beta$, $\beta_1$,$\beta_2$, $\nu$,$\nu_1$ and $\nu_2$. The different percolation phase transitions with the same critical exponents belong to the same universality class.

\section{Random site percolation on two-dimensional square lattice}
\label{Model}
On a two-dimensional square lattice, the percolation transition can be investigated both for site and bond. For the site percolation, we consider a $L\times L$ square lattices with periodic boundary conditions in both directions. At first, there are $N=L\times L$ unoccupied sites in a lattice. Then the sites in the lattice are occupied one by one under some rule. Clusters consisting of adjacent sites appear during the evolution. When $N_r$ sites are occupied in the lattice, we define a reduced number of occupied sites $r=N_r/N$. If the size of the largest cluster becomes comparable with $N$, there is a site percolation in the lattice.

For the random site percolation (SP), sites are added randomly. It is known that SP on two-dimensional square lattice occurs at $r_c=0.59276421(13)$~\cite{Newmann00}. From Coulomb gas arguments~\cite{Nienhuis} and conformal field theory~\cite{Cardy}, the critical exponents $\beta=5/36$ and $\nu=4/3$ were predicted.

In our Monte Carlo simulation of site percolation, the algorithm of Newmann and Ziff ~\cite{Newmann00,Newmann01} has been used. We have taken lattice sizes $L=32,64,128,256,512$,$1024$,$2048$,$4096$ and $8192$ in our simulations.  $12,800,000$ independent simulations have been run for each lattice size.

\begin{figure}
\resizebox{0.45\textwidth}{4cm}{ \includegraphics{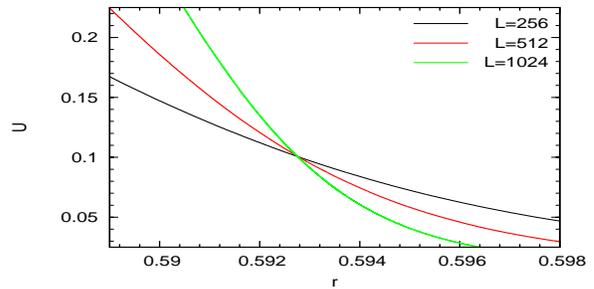} }
\caption{Size ratio of the second largest cluster to the largest cluster $U = s_2/s_1$. The critical point of site percolation $r_c=0.59276(3)$ is obtained from its cross-point.}
\label{ratio_sp}       
\end{figure}

In Fig.\ref{ratio_sp}, the size ratio $s_2/s_1$ is plotted with respect to $r$ for three different lattice sizes. From its cross-point, we get the critical point $r_c=0.59276(3)$ which agrees with the results of Ref.\cite{Newmann00}.

\begin{figure}
\resizebox{0.45\textwidth}{4cm}{ \includegraphics{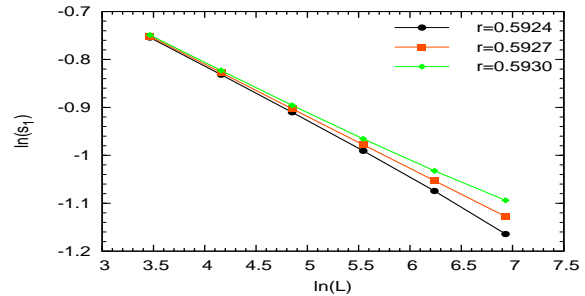} }
\caption{Log-log plot of $s_1$ with respect to $ L$ for three different reduced number of occupied sites $r$. At $r_c=0.5927(3)$, the curve becomes a straight line and its slope gives the critical exponent ratio $\beta/\nu=0.108$.}
\label{lnSMAX_sp}       
\end{figure}

\begin{figure}
\resizebox{0.45\textwidth}{4cm}{\includegraphics{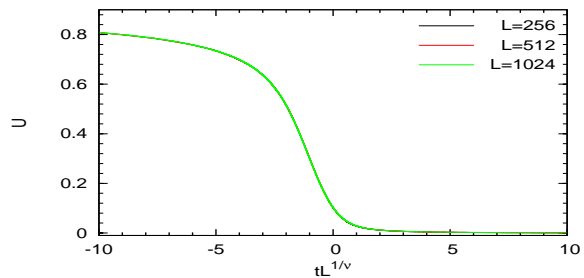} }
\caption{Finite-size scaling function $U(tL^{1/\nu})$ of the size ratio $s_2/s_1$ for random site percolation with $1/\nu=0.75$.}
\label{U_sp}       
\end{figure}

In Fig.\ref{lnSMAX_sp}, $\ln s_1$ is shown with respect to $\ln L$ for three different $r$. At $r=0.5927$, the curve is nearly a straight line with slope equal to $-0.108$. The curvature becomes negative for $r=0.5924$ and positive for $r=0.5930$. We can conclude that the critical point $r_c=0.5927(3)$ and the critical exponent ratio $\beta/\nu=0.108(4)$, which agrees well with the exact value $\beta/\nu=5/48$. Within error bounds, the critical point from $\ln s_1$ agrees with  that from $s_2/s_1$. Using $1/\nu=0.75$ in Fig.\ref{U_sp}, three curves of $s_2/s_1$ at $L=256,512,1024$ collapse into a scaling function function $U(tL^{1/\nu})$.

\begin{figure}
\resizebox{0.45\textwidth}{5cm}{ \includegraphics{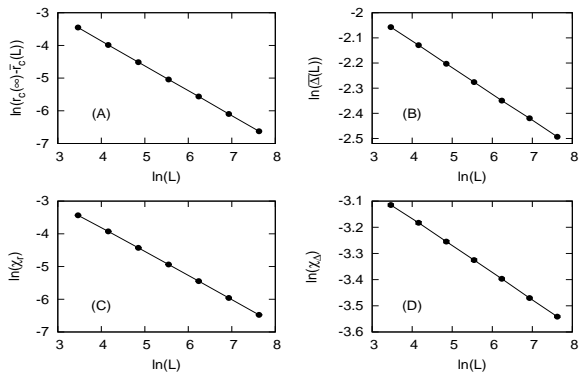} }
\caption{Critical behaviors of random site percolation during network evolution: (A)log-log plot of $r_c (\infty) - \bar{r}_c (L)$ with $r_c (\infty)=0.5927$ and the exponent $1/\nu_1=0.75(1)$; (B)log-log plot of $\bar{\Delta}(L)$ with the exponent $\beta_2=0.104(1)$; (C)log-log plot of $\chi_r$ with the exponent $1/\nu_2=0.74(1)$; (D)log-log plot of $\chi_\Delta$ with the exponent $\beta_2=0.104(1)$. }
\label{gap_sp}       
\end{figure}

In Fig.\ref{gap_sp}, we show critical behaviors of the largest cluster during the evolution of SP. The Monte Carlo simulation results of $\bar{r}_c (L)$ are demonstrated in Fig.\ref{gap_sp}(A). Using $r_c (\infty) = 0.5927(3)$ from the above investigation, a log-log plot of $ r_c (\infty) - \bar{r}_c (L) $ with respect to $L$ is made. Our Monte Carlo data demonstrate a linear dependence of  $ \ln [r_c (\infty) - \bar{r}_c (L)] $ on $\ln L$. From its slope, we obtain the inverse of exponent $1/\nu_1 = 0.75(1)$. In Fig.\ref{gap_sp}(B), the power-law finite-size behavior of $\bar{\Delta} (L)$ in Eq.(\ref{delta}) is confirmed by the Monte Carlo simulation data. The exponent $\beta_1=0.104(1)$ is calculated from the slope of the straight line.

The root mean squares of fluctuations $\chi_r$ and $\chi_\Delta$ are shown in Figs.\ref{gap_sp}(C) and (D). Our Monte Carlo simulation data of $\chi_r$ follow a power-law dependence on $L$ with the exponent $1/\nu_2=0.74(1)$. The power-law behavior of $\chi_\Delta$ in Eq.\ref{eq11} is confirm by our simulation data and we obtain $\beta_2=0.104(1)$.

We have investigated also the evolution critical behaviors of random bond percolation (BP) on two-dimensional square lattice. These results of BP are given in Table.\ref{table1}. We can see that the critical exponents of SP are equal to that of BP. Further, the critical exponent ratio $\beta/\nu$ is equal to $\beta_1$ and $\beta_2$  and $\nu$ is equal to $\nu_1$ and $\nu_2$ within error bounds.

\section{Site percolation under product rule on two-dimensional square lattice}
\label{JUMP}

In this section, we consider the site percolation under product rule (SPPR). We take the product rule ~\cite{Choi11,Bastas11} for the site evolution as follows: (1)selecting two unoccupied sites randomly; (2)calculating the product of sizes of the clusters which are connected by two chosen sites respectively; (3)the site with smaller product is occupied.

\begin{figure}
\resizebox{0.45\textwidth}{6cm}{ \includegraphics{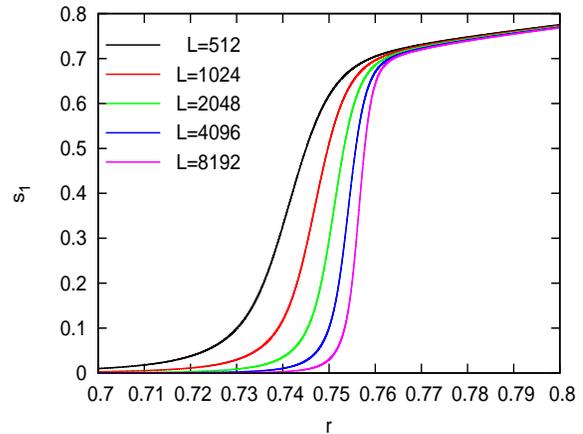} }
\caption{Reduced size of the largest cluster  as a function of $r$ at lattice size $L=512,1024,2048,4096,8192$.}
\label{fig1}
\end{figure}

The reduced sizes $s_1$ of the largest cluster for lattice size $L$ from $512$ to $8192$ are shown in Fig.\ref{fig1}. $s_1$ is nearly zero at small $r$ and becomes finite when $r$ is larger than a critical value $r_c$. Finite $s_1$ indicates the emergence of a giant cluster and a site percolation transition.

\begin{figure}
\resizebox{0.45\textwidth}{4cm}{ \includegraphics{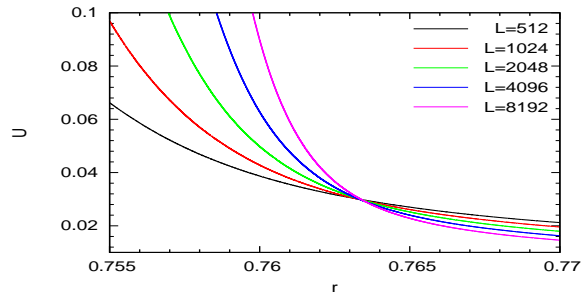} }
\caption{Size ratio of the second largest cluster to the largest cluster $U = s_2/s_1$ at different lattice sizes. Its fixed point gives the critical point $r_c=0.7634(2)$.}
\label{ratio}       
\end{figure}

\begin{figure}
\resizebox{0.45\textwidth}{4cm}{ \includegraphics{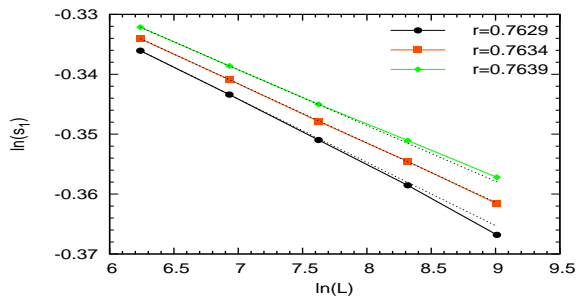} }
\caption{$\ln s_1$ with respect to $\ln L$ around the critical point $r_c=0.7634(5)$, where the curve becomes a straight line with slope equal to $-0.010$.}
\label{lnSMAX}       
\end{figure}

\begin{figure}
\resizebox{0.45\textwidth}{4cm}{\includegraphics{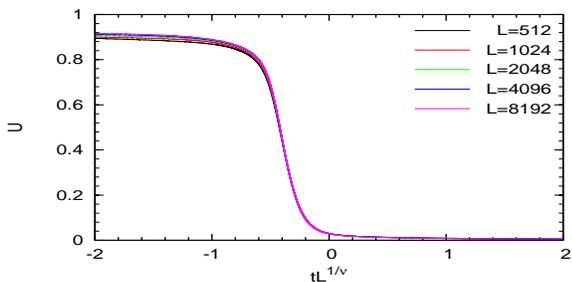} }
\caption{Finite-size scaling function $U(tL^{1/\nu})$ of size ratio $s_2/s_1$ with $1/\nu=0.42$.}
\label{U}       
\end{figure}

The results obtained from the size of the largest cluster are shown in Figs.~\ref{ratio}, \ref{lnSMAX} and \ref{U}. From the fixed point of $s_2/s_1$ in Fig.\ref{ratio}, we can get the critical point $r_c=0.7634(2)$. The straight line in Fig.\ref{lnSMAX} gives the critical point $r_c=0.7634(5)$, which is in agreement with the value of the fixed point. The accuracy of our critical point is higher than that of $r_c = 0.768(3)$ in Ref.\cite{Choi11} and $r_c = 0.756(6)$ in Ref.\cite{Bastas11}.

The slope of the straight line in Fig.\ref{lnSMAX} gives the critical exponent ratio $\beta/\nu =0.010(1)$, which is different from  $\beta/\nu=0.04\pm 0.02$ of Ref.\cite{Bastas11}.

We can see in Fig.\ref{U} that the size ratios $s_2/s_1$ of different lattice sizes collapse well with the scaling variable $tL^{1/\nu}$ for data of $L \ge 512$. This means that the asymptotic region of SPPR is more restricted than that of SP. In the finite-size scaling plot of $s_2/s_1$, we use the inverse of the critical exponent $1/\nu=0.42(1)$.

The size ratio at critical point $U(0)$ is supposed to be universal. We obtain $U(0)=0.03$ for SPPR and $U(0)=0.1$ for SP. This difference of $U(0)$ indicates also that SPPR and SP belong to different universality class.

\begin{figure}
\resizebox{0.45\textwidth}{5cm}{ \includegraphics{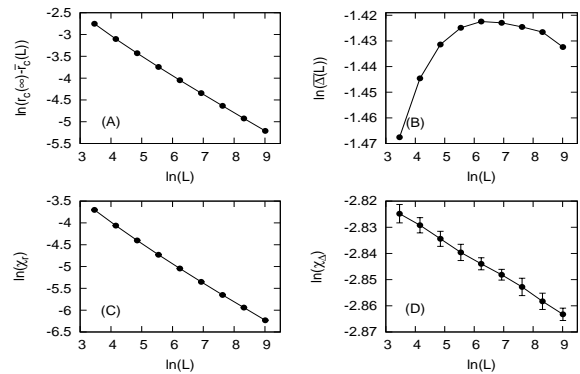} }
\caption{Results of site percolation under the product rule: (A)log-log plot of $r_c (\infty) - \bar{r}_c (L)$ with $r_c (\infty)=0.7634$ and the exponent $1/\nu_1=0.42(2)$; (B)log-log plot of $\bar{\Delta}(L)$, which increases at first and then decreases slowly; (C)log-log plot of $\chi_r$ with the exponent $1/\nu_2=0.42(3)$; (D)log-log plot of $\chi_\Delta$ with the exponent $\beta_2=0.008(4)$. }
\label{gap_sppr}       
\end{figure}

We have studied also the size jumps of the largest cluster during network evolution for SPPR. The different quantities related to the size jump are shown in Fig.\ref{gap_sppr} with respect to the lattice size $L$. From Fig.\ref{gap_sppr}(A), we can conclude that the average transition point approaches to its bulk limit according the power-law in Eq.\ref{eq8}. Using the Monte Carlo data of the lattice size $l\geq 512$, we get the inverse of exponent $1/\nu_1=0.42(2)$. The root mean square of the fluctuations of transition point is plotted in Fig.\ref{gap_sppr}(C) and  a power-law in Eq.\ref{eq10} is found $1/\nu_2=0.42(3)$. The simulation results of the size jumps of the largest cluster are presented in Figs.\ref{gap_sppr}(B) and (D). The average size jump $\bar{\Delta}$ shows a quite different $L$-dependence from that of the random site percolation. No power-law behavior is found for $\bar{\Delta} (L)$, which increases with $L$ at first and then decreases. The root mean square of fluctuations of $\Delta$ follows a power-law of the lattice size also. From Monte Carlo simulation data, we get $\beta_2=0.008(4)$.

For comparison, we have studied also the evolution critical behaviors of bond percolation under product rule (BPPR). The critical exponents related to the cluster size jump $\beta_1 = 0.044(5)$ and $\beta_2 = 0.041(4)$ are obtained. For the exponents related to critical point, we get $1/\nu_1=0.99(5)$ and $1/\nu_2=0.94(3)$. The critical exponent ratio $\beta/\nu$ and the inverse of critical exponent $1/\nu$ of BPPR have been calculated in Ref.\cite{Liu11}. $\beta/\nu = 0.064(3)$ is different from the exponents $\beta_1$ and $\beta_2$. Within error bounds, $1/\nu=0.93(1)$ is equal to $1/\nu_1$ and $1/\nu_2$. All results of BP, SP, BPPR and SPPR are summarized in Table.\ref{table1}. We can see that the critical exponents of BPPR are different from that of SPPR. The universality classes of BPPR and SPPR are different.

\begin{table}[!h]
\caption{Summary of critical points and critical exponents of BP, SP, BPPR and SPPR. $r_c$, $\beta/\nu$ and $1/\nu$ of BPPR are taken from Ref.\cite{Liu11}.}
\begin{tabular}{l p{2cm} p{2cm} p{2cm} p{2cm}}
\hline \hline & BP & SP  & BPPR & SPPR
\\\hline
$r_c$      & 0.5000(4)    & 0.5927(3)  & 0.5266(1)   & 0.7634(2)    \\
$1/\nu$	   & 0.75	  & 0.75(1)    &   0.93(1)   & 0.42(1)	\\
$1/\nu_1$ & 0.75(1)  	  &  0.75(1)   &  0.99(5)    & 0.42(2) \\
$1/\nu_2$ & 0.74(1)      &   0.74(1)   &  0.94(3)   & 0.42(3)     \\
$\beta/\nu$& 0.108(4)     &  0.108(4) &  0.064(3)    & 0.010(1)	\\
$\beta_1$ & 0.105(2)     &  0.104(1)  &  0.044(5)   &       \\
$\beta_2$ & 0.104(1)     &  0.104(1)   &  0.041(4)  & 0.008(4) \\\hline

\end{tabular}
\label{table1}
\end{table}

\section{Conclusion}
\label{Conclusion}

We have studied the critical behaviors of the sizes and the size jumps of giant clusters on two-dimensional square lattice for both site and bond percolation under random and product rules respectively. Under the random rule, sites or bonds are added randomly into the lattice. We add the site or bond with smaller size product of two connected clusters into the lattice from two candidates picked randomly when the product rule is taken. The finite-size scaling forms of giant cluster size and size jumps are proposed and used to determine the transition points and critical exponents of the percolation transitions from Monte Carlo data.

Our results show that the critical exponents of the size and size jump of random site percolation are equal to that of random bond percolation within error bounds of Monte Carlo data. As expected, the universality class of percolation is independent of site and bond under the random rule. The critical exponents of the size and size jump in the site percolation under the product rule are different from that in the random percolation and also the bond percolation under the product rule. The site percolation and bond percolation under the product rule do not belong to the same universality class.

This work is supported by the National Natural Science Foundation
of China under grant 11121403.

\end{document}